\documentstyle[12pt]{article}

   \font\tenmsb=msbm10 scaled\magstep 1
   \font\sevenmsb=msbm7 scaled \magstep 1
   \font\faivemsb=msbm5 scaled \magstep 1
\newfam\msbfam
      \textfont\msbfam=\tenmsb
      \scriptfont\msbfam=\sevenmsb
      \scriptscriptfont\msbfam=\faivemsb
\def\Bbb#1{{\fam\msbfam #1}}
\font\tengothic=eufm10 scaled\magstep 1
\font\sevengothic=eufm7 scaled\magstep 1
\newfam\gothicfam
      \textfont\gothicfam=\tengothic
      \scriptfont\gothicfam=\sevengothic

\begin{document}

\begin{center}
{\Large{\bf Self-Similar Extrapolation of Asymptotic Series and
Forecasting for Time Series} \\ [5mm]
V.I. Yukalov} \\ [3mm]
{\it Bogolubov Laboratory of Theoretical Physics \\
Joint Institute for Nuclear Research, Dubna 141980, Russia}
\end{center}

\vskip 2cm

\begin{abstract}

The method of extrapolating asymptotic series, based on the
Self-Similar Approximation Theory, is developed. Several important
questions are answered, which makes the foundation of the method
unambiguous and its application straightforward. It is shown how
the extrapolation of asymptotic series can be reformulated as
forecasting for time series. The probability measure is introduced
characterizing the ensemble of forecasted scenarios. The way of
choosing the complete family of data bases is advanced.
\end{abstract}

\newpage

\section{Introduction}

The problem of reconstructing functions from their asymptotic
series is constantly met in all branches of physics and applied
mathematics [1]. Not less ubiquitous is the problem of forecasting
for time series [2]. Here we show that these two problems have
much in common and can be solved in the same way. The solution
stems from the {\it Self-Similar Approximation Theory} [3--13].
The basics of the particular techniques to be considered here have
been suggested earlier and the feasibility of applying
them to analyzing time series has been discussed [14,15]. However
several important points remained unclear. The aim of the present
paper is to develop further the theory, so that to remove
ambiguities left, to provide a firm mathematical foundation, and
to advance some fresh ideas permitting one to extend the
applicability of the method. More concretely, I aim at providing
answers to the following questions:

\vskip 2mm

(1) What presentation of asymptotic series is the most correct and
general for their extrapolation involving self-similar exponential
approximants?

\vskip 2mm

(2) What is the result of the fractal transformation, that is, of
the multiplicative power-law transformation?

\vskip 2mm

(3) What should be the power of the fractal transformation?

\vskip 2mm

(4) How to define control functions characterizing the effective
time of motion with respect to the approximation number?

\vskip 2mm

(5) What is the correct presentation of time series, as a forward
or backward recursion, in order that forecasting for time series
could be formulated as extrapolation of asymptotic series?

\vskip 2mm

(6) How to select among several scenarios forecasted by using
different data bases?

\vskip 2mm

(7) How to construct a complete family of data bases, that is, how
many terms from the past-history data base are to be taken and how
to vary the related past-time scales?

\section{Extrapolation and Forecasting}

Let us start with the problem of extrapolating asymptotic series.
Assume, for simplicity, that the sought function $f(x)$ is real
and depends on a real variable $x$. Suppose that all we know is
the asymptotic behaviour of this function, $f(x)\simeq f_k(x)$, in
the vicinity of $x=0$. The index $k=0,1,2,\ldots$ here enumerates
approximations. The most common form of such asymptotic
approximants is the power series
\begin{equation}
\label{1}
f_k(x) = \sum_{n=0}^k \; a_n\; x^{\alpha_n} \; ,
\end{equation}
in which $a_0\neq 0$ and the powers $\alpha_n$ are arranged in the
ascending order
$$
\alpha_n < \alpha_{n+1} \qquad (n=0,1,\ldots,k) \; .
$$
One usually has $\alpha_n=n$, but, for generality, we may imply
that the powers $\alpha_n$ are arbitrary real numbers, that even
can be negative, provided the ascending order is retained.

The problem of extrapolation of asymptotic series sounds as
follows: How to define the function $f(x)$ at finite $x$ from the
knowledge of its asymptotic expansions (1) valid only in the
asymptotic vicinity of $x=0$?

In order that mathematical manipulations be independent from the
choice of physical units which the sought function is measured in,
it is more appropriate to deal with a dimensionless function. This
can be easily achieved by factorizing out the zero term $f_0(x) =
a_0x^{\alpha_0}$, with $a_0\neq 0$, and introducing the {\it
dimensionless function}
\begin{equation}
\label{2}
\varphi_k(x) \equiv \frac{f_k(x)}{f_0(x)} \; .
\end{equation}
This point is rather trivial and well understood in physics, where
one always prefers to deal with scale-invariant quantities.

Clearly, the variable $x$ is also to be dimensionless. When the
variable pertains to a finite interval, it always can be scaled so
that to be in the {\it unitary interval}
\begin{equation}
\label{3}
0\leq x \leq 1 \; .
\end{equation}
This requirement will be assumed in what follows. And it will be
shown that condition (3) is not merely nontrivial but principal.
If the initial physical variable is given on an infinite interval,
one can change the variables so that the new variable be defined
on a finite interval, and then, by the appropriate scaling, one
can reduce the latter interval to the unitary one.

Thus, we consider the asymptotic series
\begin{equation}
\label{4}
\varphi_k(x)  =\sum_{n=0}^k \; b_n\; x^{\beta_n} \; ,
\end{equation}
in which
$$
b_n \equiv \frac{a_n}{a_0} \; , \qquad \beta_n \equiv \alpha_n
-\alpha_0 \geq \beta_0 = 0 \; .
$$
The sequence $\{\varphi_k(x)\}_{k=0}^\infty$ has the zero radius
of convergence, being divergent for any finite $x$. It converges
solely at one point $x=0$. How it would be possible to extrapolate
the asymptotic expansion (4), having sense only for
$x\rightarrow 0$, to the finite interval (3)?

Such an extrapolation can be done by employing the Self-Similar
Approximation Theory [3--13]. The main idea of the latter is to
present a relation between subsequent approximants as the motion
with respect to an effective time whose role is played by the
approximation number $k$. However, it is meaningless to look for a
relation between the terms of a divergent sequence. Hence, the
first thing one has to do is to transform the divergent sequence
$\{\varphi_k(x)\}$ to a convergent form, which can be accomplished
by a transformation involving control functions [3]. The name of
the latter comes from their role in controlling convergence. In
the case of the power series (4), a convenient transformation
is the multiplicative power-law transformation
[10--12], which is simply the multiplication of $\varphi_k(x)$ by
a power-law factor $x^s$, where $s$ is a control function. Since
the power laws are generic for fractals [16], this kind of
multiplication can be termed the fractal transformation. In this
way, we introduce the {\it fractal transform}
\begin{equation}
\label{5}
\Phi_k(x,s) \equiv x^s\varphi_k(x) \; ,
\end{equation}
whose inverse is
\begin{equation}
\label{6}
\varphi_k(x) = x^{-s}\Phi_k(x,s) \; .
\end{equation}

The resulting fractal transform is
\begin{equation}
\label{7}
\Phi_k(x,s) = \sum_{n=0}^k \; b_n\; x^{s+\beta_n} \; .
\end{equation}
This can be considered as asymptotic series, for all $|x|<1$,
provided that
\begin{equation}
\label{8}
s\rightarrow \infty \; .
\end{equation}
The value of expansion (7) at $x=1$ can always be defined as the
limit from the left. Thus the sequence $\{\Phi_k(x,s)\}$ can be
treated as convergent on the whole unitary interval (3) under
condition (8).

For a convergent sequence, it becomes meaningful to try to find a
relation between subsequent terms. Such a relation, according to
the Self-Similar Approximation Theory [3-9], can be formulated as
the motion with respect to the approximation number $k$. In
mathematical parlance, the motion can be expressed as the
semigroup property for a family of endomorphisms. To this end, we
define the function $x(\varphi,s)$ by the equation
\begin{equation}
\label{9}
\Phi_0(x,s) = \varphi\; , \qquad x=x(\varphi,s) \; .
\end{equation}
Then, we introduce
\begin{equation}
\label{10}
y_k(\varphi,s) \equiv \Phi_k(x(\varphi,s),s) \; ,
\end{equation}
which is an endomorphism of $\Bbb{R}$. The semigroup property for
an endomorphism $y_k$ reads as $y_{k+p}=y_k\cdot y_p$. This is
equivalent to the property of {\it group self-similarity}
\begin{equation}
\label{11}
y_{k+p}(\varphi,s) = y_k(y_p(\varphi,s),s) \; .
\end{equation}

The family $\{ y_k|\; k=0,1,2,\ldots\}$ of the endomorphisms
$y_k$, with the semigroup property (11), forms a cascade, that is,
a dynamical system in discrete time $k$. Then Eq. (11) is the
evolution equation of the cascade, with the initial condition
\begin{equation}
\label{12}
y_0(\varphi)=\varphi \; ,
\end{equation}
which results from Eqs. (9) and (10). The functional equation (11)
does not uniquely define a solution $y_k(\varphi)$. Among the wide
class of solutions of Eq. (11) there are, in particular, the
power-law solutions $y_k(\varphi,s)=\varphi^{\alpha_k(s)}$
provided that their powers satisfy the condition $\alpha_{k+p}(s)
= \alpha_k(s) +\alpha_p(s)$. But the power-law expressions form
only a narrow class of possible solutions of Eq. (11). Hence the
group self-similarity (11) is essentially more general property
than the simple geometric self-similarity resulting in power-law
solutions [16].

To find a solution of the functional equation (11), corresponding
to a fixed point, we can proceed as follows [4--9]. Let us embed
the cascade $\{ y_k|k=0,1,2,\ldots\}$ into a flow $\{ y_\tau|\;
\tau\geq 0\}$, which implies that the latter satisfies the same
self-similarity relation (11) with the initial condition (12).
This group relation for the flow yields the Lie equation, which is
the differential equation with respect to the continuous effective
time $\tau$. The differential equation may be integrated from the
effective time $\tau=n$ to the time $n+\tau_n$ that is necessary
for reaching a quasifixed point $y_n^*$ which is an approximate
fixed point. The resulting {\it evolution integral} is
\begin{equation}
\label{13}
\int_{y_n}^{y_n^*}\; \frac{d\varphi}{v_n(\varphi,s)} = \tau_n\; ,
\end{equation}
where
\begin{equation}
\label{14}
v_n(\varphi,s) \equiv y_n(\varphi,s)-y_{n-1}(\varphi,s) =
b_n\; \varphi^{1+\beta_n/s}
\end{equation}
is the cascade velocity. From the evolution integral (13), we get
the quasifixed point $y_n^*=y_n^*(\varphi,s)$, which, according to
Eqs. (9) and (10), defines
\begin{equation}
\label{15}
\Phi^*_k(x,s) \equiv y_k^*(x^s,s) \; .
\end{equation}
Then we have to realize the inverse fractal transformation (6),
with the limit (8), which gives
\begin{equation}
\label{16}
\varphi^*_k(x) \equiv \lim_{s\rightarrow\infty}
x^{-s}\;\Phi_k^*(x,s) \; .
\end{equation}
This procedure is to be accomplished so many times as necessary
for reorganizing all series entering $\varphi_k^*(x)$. Such a
repeated procedure has been called [11] self-similar bootstrap.
Here we suggest the fastest way of accomplishing this bootstrap,
when the $k$-order series $\varphi_k(x)$ requires just $k$
iterations. To this end, we may present the asymptotic expansion
(4) as
\begin{equation}
\label{17}
\varphi_k(x) \equiv 1 + x_1 \; ,
\end{equation}
with the sequence of the iterative terms
$$
x_1 =\frac{b_1}{b_0}\; x^{\beta_1-\beta_0}\left ( 1 + x_2\right )\; ,
\qquad x_2 =\frac{b_2}{b_1}\; x^{\beta_2-\beta_1}\left ( 1 +
x_3\right ) \; , \qquad \ldots
$$
\begin{equation}
\label{18}
x_n=\frac{b_n}{b_{n-1}}\; x^{\beta_n-\beta_{n-1}}(1 +
x_{n+1})\; , \qquad \ldots,\qquad  x_k =\frac{b_k}{b_{k-1}} \;
x^{\beta_k-\beta_{k-1}} \; .
\end{equation}
Then we may consider each expression $1+x_n$ as asymptotic series
with respect to small $x_n\rightarrow 0$. The self-similar
renormalization described above results in transforming $1+x_n$ to
$\exp(x_n\tau_n)$. Starting with the form (17), we accomplish $k$
steps of this self-similar interactive procedure. Then, with the
notation
\begin{equation}
\label{19}
c_n \equiv \frac{a_n}{a_{n-1}}\; \tau_n\; , \qquad
\nu_n \equiv \alpha_n -\alpha_{n-1} \; ,
\end{equation}
where $n=1,2,\ldots,k$, we come to the {\it exponential
self-similar approximant}
\begin{equation}
\label{20}
\varphi_k^*(x) =\exp\left ( c_1x^{\nu_1}\; \exp\left (
c_2x^{\nu_2} \ldots\exp\left ( c_kx^{\nu_k}\right )\right )\ldots
\right ) \; .
\end{equation}

What is yet left undefined in the approximant (20) is the
effective times, $\tau_n$, which play the role of control
functions. The latter may be determined from fixed-point
conditions [3,12]. Here we advance a more general way of defining
control functions $\tau_n$. This way is common for optimal control
theory, where control functions are defined from the minimization
of a cost functional. To construct the latter, one has to
formulate the requirements imposed on the sought control
functions. In our case, we would like to reach a quasifixed point
as fast as possible. The minimal effective time for this
corresponds to one step of the iterative procedure. If we reach an
answer in $n$ steps, then the minimal effective time $\tau_n$
should be close to $1/n$. Another requirement characteristic of
the motion near a quasifixed point is the smallness of the
distance $v_n\tau_n$ passed at the $n$-th step. This suggests to
construct the {\it time-distance cost functional}
\begin{equation}
\label{21}
F =\frac{1}{2}\; \sum_n
\left [ \left ( \tau_n -\; \frac{1}{n}\right )^2 +
(v_n\tau_n)^2\right ] \; .
\end{equation}
Minimizing $F$ with respect to $\tau_n$ yields
\begin{equation}
\label{22}
\tau_n =\frac{1}{n(1+v_n^2)} \; .
\end{equation}
Here the $x$-representation of the cascade velocity $v_n(x)\equiv
x^{-s}v_n(x^s,s)$ results in
\begin{equation}
\label{23}
v_n(x) = \varphi_n(x) - \varphi_{n-1}(x) = b_nx^{\beta_n} \; .
\end{equation}
Summarizing, for expressions (19) we have
$$
c_n(x) =\frac{a_n}{a_{n-1}} \; \tau_n(x) \; , \qquad
\nu_n=\alpha_n -\alpha_{n-1} \; ,
$$
\begin{equation}
\label{24}
\tau_n(x) =\frac{1}{n[1+v_n^2(x)]}\; , \qquad
v_n(x) =\frac{a_n}{a_0}\; x^{\alpha_n-\alpha_0} \; .
\end{equation}
The functions $c_n(x)$, being proportional to the control
functions $\tau_n(x)$, can be called {\it controllers}.

Finally, for the asymptotic series (1), we obtain the {\it
self-similar extrapolation}
\begin{equation}
\label{25}
f_k^*(x) = f_0(x)\exp\left ( c_1x^{\nu_1}\; \exp\left (
c_2x^{\nu_2}\ldots \exp\left ( c_kx^{\nu_k}\right )\right ) \ldots
\right ) \; ,
\end{equation}
where $c_n=c_n(x)$ and $\nu_n$ are given in Eq. (24), and which
extrapolates the series (1), valid only for $x\rightarrow 0$, to
the whole interval (3).

Now we shall show how the extrapolation of asymptotic series can
be reformulated as forecasting for time series. This
reformulation, as will be evident from what follows, requires that
the considered time series be presented as a {\it backward
recursion}, with the related moments of time being arranged in the
{\it backward order}
\begin{equation}
\label{26}
t_{n+1} < t_n \qquad (n=0,1,2,\ldots,k) \; .
\end{equation}

The time series for a quantity of interest is an ordered set
$\{ f_n\}$ of the values $f_n$ of this quantity, measured at the
corresponding times $t_n$. The past time horizon is the time
interval $[t_k,0]$ when the series is observed. The present time
is chosen to be at $t_0=0$. Let the considered time series be
described by the past-history {\it data base}
\begin{equation}
\label{27}
\Bbb{D}_k \equiv \{ f_k,f_{k-1},\ldots,f_0|\; t_k<t_{k-1}< \ldots
<0\} \; .
\end{equation}
The problem to be attacked is how, on the grounds of the given
data, to predict the behaviour of the time series in future times
$t>0$?

In order to employ the technique developed for asymptotic series,
we need, first, to construct a sequence of functions approximating
the behaviour of the time series at asymptotically small time
$t\rightarrow 0$. Similarly to the case of asymptotic series, we
have to work with the quantities reduced to the scale-invariant
dimensionless form. And the time variable is to be normalized with
respect to the length of the {\it prediction horizon} so that this
be a unitary interval
\begin{equation}
\label{28}
0\leq t\leq 1\; .
\end{equation}

For the given data base (27), one may construct an interpolating
function $f_k(t)$ such that
\begin{equation}
\label{29}
f_k(t_n) = f_n \qquad (n=0,1,\ldots,k) \; .
\end{equation}
This interpolation can be uniquely realized e.g. by means of the
Lagrange interpolation formula [17] as the sum
\begin{equation}
\label{30}
f_k(t) =\sum_{n=0}^k\; f_n\; l_n^k(t) \qquad (k\geq 1)
\end{equation}
over the Lagrange polynomials
$$
l_n^k(t) \equiv \prod_{m(\neq n)}^k \frac{t-t_m}{t_n-t_m} \qquad
(n\leq k) \; .
$$
Formula (30) can be rewritten as the power series
\begin{equation}
\label{31}
f_k(t) = \sum_{n=0}^k a_n\; t^n \; .
\end{equation}
The coefficients $a_n$ depend on the given data base (27), but an
additional index $k$ is not explicitly shown here for brevity.
Expressions (30) or (31), by construction, satisfy Eq. (29),
including the moment of time $t_0=0$, where
\begin{equation}
\label{32}
f_k(0) = f_0 \; .
\end{equation}
Therefore, the interpolative formula (31) can be considered as an
asymptotic series with respect to $t\rightarrow +0$.

In the same way as asymptotic series, expression (31) has no sense
for finite $t>0$, that is, cannot be directly employed for
predicting future. But, being just a particular kind of asymptotic
series, Eq. (31) can be extrapolated to the whole interval (28)
following the general method of self-similar extrapolation. Then,
analogously to Eq. (25), we obtain the {\it self-similar forecast}
\begin{equation}
\label{33}
f_k^*(t) = f_0\;\exp\left ( c_1 t\; \exp\left (
c_2 t\ldots \exp\left ( c_k t\right )\right ) \ldots
\right ) \; ,
\end{equation}
where $c_n=c_n(t)$ are the {\it controllers}
$$
c_n(t) =\frac{a_n}{a_{n-1}}\; \tau_n(t) \qquad
(n=1,2,\ldots,k) \;,
$$
\begin{equation}
\label{34}
\tau_n(t) =\frac{1}{n[1+v_n^2(t)]} \; \qquad
v_n(t) =\frac{a_n}{f_0}\; t^n \; .
\end{equation}

In this way, for each given past-history data base (27), we may
construct the forecast (33) valid for the future time horizon
(28). Since the coefficients $a_n\equiv a_{nk}$ depend on the
choice of the data base, according to the conditions (29),
the controllers $c_n(t) \equiv c_{nk}(t)$ also depend on
the data base. This means that, for another data base, we shall
obtain another value of the forecast. There are two things that
may be varied for data bases: the number of points $k$ and the set
$\{ t_n\}$ of the time moments. Denoting a particular choice of
the latter set with the index $j$, we have $\{ t_n^{(j)}\}$. Thus,
each data base has to be labelled by two indices
\begin{equation}
\label{35}
\Bbb{D}_k(j) \equiv \{ f_k^{(j)},f_{k-1}^{(j)},\ldots,f_0|\;
t_k^{(j)} < t_{k-1}^{(j)} < \ldots < 0\} \; .
\end{equation}
Varying these indices produces a family $\{\Bbb{D}_k(j)\}$ of the
data bases and, respectively, results in an ensemble
$\{ f_k^*(j,t)\}$ of the self-similar forecasts. How could we
classify this ensemble of possible scenarios?

The problem of classifying scenarios in forecasting is analogous
to the problem of pattern selection in the case of nonunique
solutions of nonlinear partial differential equations [18]. A
probabilistic approach to the problem of pattern selection has
been recently advanced [19], which can also be used for defining a
probability measure for the ensemble of forecasted scenarios.
Following the approach [19], we treat the map $\{ f_k^*(j,t)|\;
k=0,1,2,\ldots\}$ as the image of a dynamical cascade, where $k$
plays the role of discrete time, and for the probability of a
forecast $f_k^*(j,t)$ we have
\begin{equation}
\label{36}
p_k(j,t) =\frac{1}{Z_k(t)}\;\exp\left\{ - \Delta S_k(j,t)
\right \} \; ,
\end{equation}
where $Z_k(t)$ is a normalization factor and $\Delta S_k(j,t)$ is
an entropy variation. Let us introduce the mapping multiplier
\begin{equation}
\label{37}
m_k(j,t) \equiv \frac{\delta f_k^*(j,t)}{\delta f_1^*(j,t)} =
\frac{\partial f_k^*(j,t)/\partial t}
{\partial f_1^*(j,t)/\partial t}
\end{equation}
and the average multiplier $\overline m_k(t)$, such that
\begin{equation}
\label{38}
\frac{1}{|\overline m_k(t)|} \equiv \sum_j\;
\frac{1}{|m_k(j,t)|} \; .
\end{equation}
Then the {\it scenario probability} (36) can be reduced [19] to the
form
\begin{equation}
\label{39}
p_k(j,t) =\left | \frac{\overline m_k(t)}{m_k(j,t)}\right | \; .
\end{equation}
The most probable scenario is, by definition, that one
corresponding to the maximal probability (39).

It may happen that the data-base variation is limited by the
information available. Then there is no choice as to deal with the
given family $\{\Bbb{D}_k(j)\}$. But if this variation can be done
in  a very wide range, the question remains as when should one
stop varying data bases? A brief answer to this question is: Vary
data bases until reaching numerical convergence. More precisely,
the variation procedure can be realized as follows. Let us fix a
time scale, labelled by $j$, and increase the data-base order
$k$, that is the number of points from the past history. The
increase of $k$ is to be stopped at the {\it saturation number}
$N_j$, after which the forecasts $f_k^*(j,t)$, for $k\geq N_j$,
practically (within a given accuracy) do not change. Then the
forecasts (33) as well as the scenario probability (39) become
dependent only on the time scale, labelled by $j$,
\begin{equation}
\label{40}
f^*(j,t) \equiv \lim_{k\rightarrow N_j} f_k^*(j,t) \; , \qquad
p(j,t) \equiv \lim_{k\rightarrow N_j} p_k(j,t) \; .
\end{equation}

Varying the data-base time scale, it looks convenient to deal with
equidistant steps $\Delta_j\equiv t_n^{(j)} - t_{n+1}^{(j)}$,
where $n=0,1,\ldots,N_j$, although the equal distance between the
time moments is, in general, not obligatory. It is natural to
start with the step $\Delta_j=1$, coinciding with the prediction
horizon. Then we may increase as well as decrease the time grid,
which can be done in different ways, say by following the law
$\Delta_{2j}=2^{-j},\; \Delta_{2j+1}=2^j$, with $j=0,1,2,\ldots$.
Again, the further variation of the time mesh is to be stopped at
$j=j_{max}$ as soon as we reach numerical convergence. In this
way, the {\it complete data family} is a set
$\{\Bbb{D}_{N_j}(j)|\; j=0,1,2,\ldots,j_{max}\}$ corresponding to
the ensemble $\{ f^*(j,t)| j=0,1,\ldots,j_{max}\}$ that can be
called the {\it scenario spectrum}. As far as this ensemble of
forecasts is weighted with the probability measure $p(j,t)$, it is
straightforward to define the expected forecast, the dispersion,
and other quantities typical of probabilistic ensembles.

\section{Conclusion}

The possibility of forecasting the main trends of time series is
of great importance and can have various applications. One of such
exciting application would be with respect to economics and
financial time series. The behaviour of markets is nowadays studied
by different methods imported from physics [20]. However the
problem of quantitative prediction for markets remains yet a
puzzle. I would think that the novel approach, developed in this
paper, is a practical step for solving the problem of forecasting
for time series, including financial time series. The aim of this
communication has been to present a mathematical foundation for
the approach, whose applications will be considered in separate
publications. The content of this paper answers the questions
posed in the Introduction, and these answers are:

\vskip 2mm

(1) The most general and correct presentation of asymptotic
series, appropriate for the self-similar extrapolation is in a
dimensionless scale-invariant form, but with the domain of a
variable reduced to the unitary interval.

\vskip 2mm

(2) The fractal transformation extends the region of validity of
the initial asymptotic series.

\vskip 2mm

(3) The power of the fractal transformation must asymptotically
tend to infinity in order to yield series with the unitary radius
of convergence.

\vskip 2mm

(4) The effective time of motion with respect to the approximation
number is to be defined from the minimum of a cost functional.

\vskip 2mm

(5) In order that extrapolation of asymptotic series could be
correctly reformulated as forecasting for time series, the latter
are to be presented as a backward recursion.

\vskip 2mm

(6) All forecasted scenarios, derived by employing different data
bases, compose a statistical ensemble equipped with a probability
measure characterizing the probabilities of particular patterns.

\vskip 2mm

(7) The complete data family is a set of data bases, with time
scales varying in the range limited by the saturation of
calculations, and with the numbers of past-history terms
sufficient for providing numerical convergence for each given time
grid.

\vskip 5mm

{\bf Acknowledgement}

\vskip 2mm

I am grateful to E. Yukalova for discussions and advice.

\end{document}